\documentclass[conference,anonymous]{IEEEtran}

\usepackage[left=0.95in, right=0.95in, top=0.75in, bottom=1.15in]{geometry}

\usepackage{booktabs} 
\usepackage[table,xcdraw]{xcolor}
\usepackage{sidecap}
\usepackage{booktabs} 
\usepackage{url}
\usepackage{subfigure}
\usepackage{mathrsfs}  
\usepackage{bm}

\usepackage{colortbl}
\usepackage[numbers]{natbib}
\usepackage[utf8]{inputenc}
\usepackage{enumitem}
\usepackage{graphicx}
\usepackage{amsmath}
\usepackage{amssymb}
\usepackage{multirow}
\usepackage{epstopdf} 
\usepackage{float}
\usepackage[defaultcolor=red]{changes} 

\newcolumntype{M}[1]{>{\centering\arraybackslash}m{#1}}
\usepackage{amsmath,epsfig,psfrag} 
\usepackage{amsfonts,color,graphics,graphicx}
\usepackage{caption}
\usepackage{algpseudocode}
\usepackage{multirow}
\usepackage{rotating}
\usepackage{multirow}
\usepackage{tabularx}

\usepackage[normalem]{ulem}
\renewcommand{\baselinestretch}{1}

\usepackage{xcolor}

\usepackage{verbatim}
\usepackage{tikz,pgfplots}
\usepackage{pgfplotstable}
\newcolumntype{P}[1]{>{\centering\arraybackslash}p{#1}}
\pgfplotsset{compat=1.15}
\setlist[itemize]{leftmargin=*}
\setlist[enumerate]{leftmargin=*}

\newcommand{\jhc}{\textcolor{blue}}

\usepackage{algorithm} 
\def\BibTeX{{\rm B\kern-.05em{\sc i\kern-.025em b}\kern-.08em
    T\kern-.1667em\lower.7ex\hbox{E}\kern-.125emX}}
\begin{document}

\title{Winning the Social Media Influence Battle: Uncertainty-Aware Opinions to Understand and Spread True Information via Competitive Influence Maximization
}
\author{
\IEEEauthorblockN{Qi Zhang}
\IEEEauthorblockA{\textit{Department of Computer Science}\\
\textit{Virginia Tech}\\
Falls Church VA, USA \\
qiz21@vt.edu}
\and
\IEEEauthorblockN{Lance M. Kaplan}
\IEEEauthorblockA{\textit{US DEVCOM Army Research Laboratory} \\ 
Adelphi MD, USA\\
lance.m.kaplan.civ@army.mil}
\and
\IEEEauthorblockN{Audun J{\o}sang}
\IEEEauthorblockA{\textit{Department of Informatics}\\
\textit{University of Oslo}\\
Oslo, Norway \\
josang@ifi.uio.no}
\and
\IEEEauthorblockN{Dong Hyun Jeong}
\IEEEauthorblockA{\textit{Department of Computer Science}\\
\textit{University of the District of Columbia}\\
Washington DC, USA \\
djeong@udc.edu}
\and
\IEEEauthorblockN{Feng Chen}
\IEEEauthorblockA{\textit{Department of Computer Science}\\
\textit{University of Texas at Dallas}\\
Richardson TX, USA \\
feng.chen@utdallas.edu}
\and
\IEEEauthorblockN{Jin-Hee Cho}
\IEEEauthorblockA{\textit{Department of Computer Science}\\
\textit{Virginia Tech}\\
Falls Church VA, USA \\
jicho@vt.edu}
}
\maketitle

\begin{abstract}
Competitive Influence Maximization (CIM) involves entities competing to maximize influence in online social networks (OSNs). Current Deep Reinforcement Learning (DRL) methods in CIM rely on simplistic binary opinion models (i.e., an opinion is represented by either 0 or 1) and often overlook the complexity of users' behavioral characteristics and their prior knowledge. We propose a novel DRL-based framework that enhances CIM analysis by integrating Subjective Logic (SL) to accommodate uncertain opinions, users' behaviors, and their preferences. This approach targets the mitigation of false information by effectively propagating true information. By modeling two competitive agents, one spreading true information and the other spreading false information, we capture the strategic interplay essential to CIM. Our framework utilizes an uncertainty-based opinion model (UOM) to assess the impact on information quality in OSNs, emphasizing the importance of user behavior alongside network topology in selecting influential seed nodes. Extensive experiments demonstrate that our approach significantly outperforms state-of-the-art methods, achieving faster and more influential results (i.e., outperforming over 20\%) under realistic network conditions. Moreover, our method shows robust performance in partially observable networks, effectively doubling the performance when users are predisposed to disbelieve true information.
\end{abstract}    	

\begin{IEEEkeywords}
Competitive influence maximization, deep reinforcement learning, uncertainty, opinion models, influence propagation
\end{IEEEkeywords}
\vspace{-2mm}
\section{Introduction} \label{sec:intro} 

Online Social Networks (OSNs) are primary platforms for information exchange and opinion formation. In situations where exclusive decisions are required, such as service selection or voting, a competitive environment arises. This sets the stage for a Competitive Influence Maximization (CIM) problem, where political parties and corporations vie to sway these decisions. CIM involves these entities strategically selecting key individuals, called seed nodes in a network, within the network to act as opinion leaders and maximize the spread of their influence throughout the network.  This study tackles the CIM problem in OSNs, focusing on two opposing parties: the {\em true party} disseminating true information and the {\em false party} spreading false information. 

Our work explores strategies for the true party to effectively counteract the false party and curb the spread of false information, given its harmful consequences, such as damaged reputations, financial losses, and manipulated public opinion~\cite{guo2020online}. The aim is to enhance the propagation of true information and mitigate the negative impacts of false information in the OSN.

Prior studies on CIM in OSNs~\cite{Bharathi07-CIM, Chung2019-DeepRL, Lin15_LearningBasedCIM, Pham19-CIM} typically represented user opinions as fixed and binary, ignoring the complex, dynamic nature of individual preferences and behaviors. They mainly leveraged the Linear Threshold (LT) model and relied heavily on network topology, resulting in inconsistent effectiveness across different network topologies~\cite{Lin15_LearningBasedCIM}. This explains the need for adaptable strategies to varying network structures and evolving competitive dynamics. Reinforcement Learning (RL) emerges as a viable approach, facilitating real-time data gathering and strategy optimization to address the intricate requirements of CIM tasks.

We present a Deep RL (DRL)-based CIM framework to enhance the dissemination of accurate information and counter false information. We aim to identify an optimal set of seed users to maximize information spread, acknowledging that user opinions are not simply binary but are formed and evolve over time. To this end, we made the following {\bf key contribution} in our work: 
\begin{enumerate}
\item {\bf Refining CIM Analysis with Uncertain Opinions in OSNs}: Traditional CIM research often reduces user opinions to binary states, inadequately reflecting the dynamic shifts in user opinions within OSNs~\cite{Ali20-CIM, Ali21_DRLCIM, Chung2019-DeepRL, Lin15_LearningBasedCIM}. We incorporate {\em Subjective Logic} (SL)~\cite{Josang16} into our CIM framework to accurately model the uncertain and subjective nature of user beliefs. This approach provides a deeper understanding of opinion evolution in OSNs, enhancing CIM analysis.
\item {\bf Enhancing CIM with Dual DRL Agents}: Previous CIM research utilizing DRL often focused on a single influence strategy and simplified user opinions to binary states, missing real-world complexities~\cite{Chung2019-DeepRL, li2020community, Lin15_LearningBasedCIM}. We present a dual-agent DRL framework that models the dissemination of both true and false information. This approach considers network structures and user behaviors, providing a comprehensive method for selecting influential seed nodes and deepening the strategic dynamics within CIM.
\item {\bf Evaluating UOM's Impact on OSN Information Quality}: We explore the effectiveness of the Uncertainty-aware Opinion Model (UOM) in enhancing information quality within OSNs. Our findings reveal that UOM captures the dynamics of false information, which often dominates before the truth emerges, posing challenges in correcting established misconceptions. Implementing UOM improved user engagement with content critically, reducing the spread and impact of false information. Further, these results affirm the UOM's potential to enhance information veracity in OSNs.

\item {\bf Exploring CIM Performance under Partial Observability with UOM}: While the impact of partially observable networks (PONs) on CIM has been recognized~\cite{lin2015analyzing, nasim2016PartialObservable}, the influence of non-binary opinions modeled by specific opinion frameworks like UOM on CIM's observability remains underexplored. Our work aims to delve into how partial observability impacts CIM performance using the UOM framework with non-binary opinion models.

\item {\bf Identifying Traits for Influential Seed Node Selection}: Through simulations of various scenarios with two parties employing different strategies, we identify critical user behavioral traits and network characteristics that optimally indicate influential seed nodes.
\end{enumerate}

We specifically use the term {\em false information} instead of {\em misinformation} or {\em disinformation} to maintain clarity. Misinformation refers to inaccurately shared information without intent to deceive, stemming from misunderstanding or ignorance. In contrast, disinformation is deliberately spread to mislead others, characterized by malicious intent. The nuanced differences between misinformation and disinformation, along with their separate impacts, fall outside the purview of our research focus.

\section{Related Work}\label{sec:related-work}

The Influence Maximization (IM) problem, initially conceptualized in~\citet{domingos2001IM} as an algorithmic challenge, involves maximizing influence across OSNs by selecting optimal seed sets. \citet{Kempe-03} further refined it into a discrete stochastic optimization problem, enhancing its theoretical framework. CIM extends the IM problem, featuring multiple entities competing to maximize their influence, with various methodologies demonstrating effectiveness in addressing CIM challenges.

\subsection{Propagation Models} 
Propagation models like Independent Cascade (IC)~\cite{kempe2003maximizing} and Linear Threshold (LT)~\cite{granovetter1978threshold} are established frameworks for analyzing network influence~\cite{Bharathi07-CIM, Kahr-2021, liu2020CIM_UnwanedUser, Pham19-CIM}. The IC model activates a node and attempts independent influence on its neighbors, whereas the LT model activates a node if the cumulative influence from neighbors surpasses a preset threshold.

Despite their wide acceptance, both models are limited in real-world scenarios due to static probabilities and thresholds, which overlook individual response variations to information. Our study enhances these models by integrating diverse opinion dynamics, particularly evaluating their effectiveness against the spread of false information.

\subsection{DRL-based CIM}

\citet{Lin15_LearningBasedCIM} introduced {\tt STORM}, a \underline{ST}rategy-\underline{O}riented \underline{R}einforce\underline{M}ent-Learning-based framework for multi-round CIM within the competitive LT model~\cite{shakarian2015IC&LT}, marking the first use of RL in CIM. {\tt STORM} employed RL to dynamically select IM strategies based on evolving user opinions and competitor actions. Building on this, \citet{Chung2019-DeepRL} developed a DRL-based CIM framework, enhancing strategic seed selection across multiple rounds using Deep-Q Learning (DQN) and a spectral community detection method to optimize seed choices and improve influence spread. Furthermore, \citet{Ali20-CIM} extended DRL applications to CIM in unknown network topologies, adapting STORM's reward and action spaces with additional network exploration actions.

However, the prior work in~\cite{Chung2019-DeepRL, Lin15_LearningBasedCIM} often assumed complete knowledge of network topology, an unrealistic scenario in practical applications. Moreover, DRL approaches in~\cite{Ali20-CIM, Ali21_DRLCIM} simplified opinion evolution with binary models. Our work addresses these limitations by integrating detailed opinion dynamics and accounting for uncertainties in opinions and network structure, thus developing a more realistic CIM framework.

\section{Problem Statement} \label{sec:problem-statement}
An OSN is modeled as an unweighted, undirected graph $G = (V, E)$, with $V$ as users and $E$ as relationships. Within $G$, two parties, $\mathcal{T}$ for true information and $\mathcal{F}$ for false information, aim to maximize their influence. Their central decision makers (CDMs) select seed users $\mathbf{S}^{\mathcal{T}}$ and $\mathbf{S}^{\mathcal{F}}$ to spread their respective information types. DRL is employed to strategically select action $\bm{A}$ to select seed nodes for maximizing influence.  Each party's node selection process will follow the procedures and opinion models described in Section~\ref{sec:our-algorithm}.  

To formally put, each party's CDM seeks to optimize the objective of maximizing their influence by:
\begin{eqnarray}
\label{eq:objective-function}
\arg \max_{\bm{S}, \bm{A}} \; \; \mathcal{IF} (G, \bm{S}, {\bm{uc}}),
\end{eqnarray}
where ${\bm{uc}}$ represents the characteristics of users, detailed in Section~\ref{subsec:opinion-models}.

\section{Proposed CIM Framework} \label{sec:our-algorithm}

\subsection{SL-based Opinion Formulation} \label{subsec:sl-op-form}
In an OSN, users adopt opinions modeled by {\em Subjective Logic (SL)}, accommodating multidimensional uncertainties~\cite{Josang16, Josang18-fusion}. SL articulates a binomial opinion $\omega = (b, d, u, a)$, incorporating belief ($b$), disbelief ($d$), uncertainty ($u$), and base rate ($a$), where $b + d + u = 1$ and each element is ranged within $[0, 1]$ as a real number. Here, $b$ reflects agreement with true information, $d$ captures disagreement with false information (or disbelieving true information), $u$ represents uncertainty due to insufficient evidence, and $a$ (and $1-a$) indicates the prior belief favoring true information (i.e., belief) or false information (i.e., disbelief).  A user's opinion update adjusts these elements based on evidence for or against belief and disbelief. An opinion $\omega$ is formulated by:
\begin{eqnarray} 
\label{eq:sl-mapping}
b = \frac{r}{r+s+W}~,
d = \frac{s}{r+s+W}~,
u = \frac{W}{r+s+W}~,
\end{eqnarray}
where $r$ and $s$ are the numbers of evidence to support $b$ and $d$, respectively, and $W$ refers to the number of uncertain evidence that cannot be judged as true or false, supporting neither $b$ nor $d$.

We denote user $i$'s opinion by $\omega_i = (b_i, d_i, u_i, \bm{a}_i)$. Since users make decisions from the two choices, $b_i$ or $d_i$, they interpret uncertainty based on their prior belief, $\bm{a}_i = \{a_i, 1-a_i\}$ to support $b_i$ and $d_i$, respectively, and $a_i + (1-a_i)=1$.  Users incorporate the projected belief or disbelief and uncertainty to make decisions in practice.  The projected belief $P(b_i)$ and disbelief $P(d_i)$ are obtained by:
\begin{eqnarray} 
\label{eq:expected-opinion}
P(b_i) = b_i + a_i \times u_i, \;
P(d_i) = d_i + (1-a_i) \times u_i,
\end{eqnarray}
where $P(b_i)+P(d_i)=1$ and $a_i u_i + (1-a_i) u_i =u_i$ with $b_i + d_i + u_i = 1$. 

We consider two types of uncertainty in user $i$'s opinion~\cite{Josang18-fusion}: vacuity and dissonance. {\em Vacuity} refers to uncertainty caused by a lack of evidence.  {\em Dissonance} indicates uncertainty due to conflicting evidence. Vacuity uncertainty mass in user $i$'s opinion, $\omega_i$, is measured by $u_i$. Dissonance uncertainty mass is defined as:
\begin{gather}
\bm{b}_i^{\mathrm{Diss}} = {(b_i + d_i) \cdot \mbox{Bal}(b_i,d_i)}, 
\label{eq:belief-dissonance}
\end{gather}
where $\bm{b}_i(x) = \{b_i, d_i\}$ and the relative mass balance between belief masses, $b_i$ and $d_i$, is given by:
\begin{equation}
\label{eq:belief-balance}
\mbox{Bal}(b_i,d_i) = 1 - \frac{|b_i-d_i|}{b_i+d_i}.
\end{equation}
We incorporated uncertainty estimates, such as vacuity and dissonance, to develop a UOM in Section~\ref{subsec:opinion-models}.

\subsection{User Types} \label{subsec:user-types}
We consider three types of users as follows:
\begin{itemize}
\item {\em True information propagators (TIPs)} have their opinion initialized by $\omega_T = (b, d, u, a) = (b \rightarrow 1, d \rightarrow 0, u \rightarrow 0, a=1)$. This opinion implies TIPs have a strong belief in true information ($b$ is close to 1), while they lack belief in false information ($d$ is close to 0).  TIPs selected as seed nodes will not change their opinions.
\item {\em False information propagators (FIPs)} have their opinion, initialized by $\omega_F=(b, d, u, a) = (b \rightarrow 0, d \rightarrow 1, u \rightarrow 0, a=0)$. This means FIPs have a strong belief in false information ($d$ is close to 1) while having a lack of belief in true information ($b$ is close to 0). FIPs selected as seed nodes will not change their opinions.
\item {\em Legitimate users} have a highly uncertain opinion, initialized by $\omega_L=(b, d, u, a) = (b \rightarrow 0, d \rightarrow 0, u \rightarrow 1, {\bm a})$. The users' prior beliefs are initialized based on uniform distribution with $a_i = 1-a_i = 0.5$ and will be updated based on the consensus operation in Eq.~\eqref{eq:consensus_op}. 
\end{itemize}

\subsection{Opinion Models} \label{subsec:opinion-models}

Each user $i$'s behavior is characterized by the following main components: opinion updating, sharing, and reading, denoted by a set of user behaviors, $\bm{uc}_i$. In our objective function, described in Eq.~\eqref{eq:objective-function}, $\bm{uc}$ refers to a set of users about these behavioral components. To formally put, $\bm{uc}$ is defined by: 
\begin{eqnarray}
\bm{uc} = \{\bm{uc}_1, \ldots, \bm{uc}_i, \ldots, \bm{uc}_n\}, 
\end{eqnarray}
where $\bm{uc}_i = \{\mathrm{updating}_i, \mathrm{reading}_i, \mathrm{sharing}_i\}$. Details about each behavioral component are shown below.

\subsubsection{\bf Opinion Updating} \label{susbsubsec:opinion-updating} 

Recall that we choose SL as it offers the ability to formulate a subjective opinion that can handle multiple types of uncertainty caused by different causes (i.e., vacuity or dissonance).  At time $t$, when user $i$ encounters user $j$ and reads user $j$'s information (i.e., $\omega_j(t)$), user $i$ will update the opinion as $\omega_i \oplus \omega_{i \otimes j} = (b_i \oplus b_{i \otimes j}, d_i \oplus d_{i \otimes j}, u_i \oplus u_{i \otimes j}, a_i \oplus a_{i \otimes j})$, and $\omega_{i \otimes j}$ indicates user $j$'s opinion based on user $i$'s trust in user $j$.  The $\omega_{i \otimes j} = (b_{i \otimes j}, d_{i \otimes j}, u_{i \otimes j}, a_{i \otimes j})$ is obtained where each component of the opinion is estimated by: 
\begin{gather} \label{eq:discounting}
b_{i \otimes j}  =  c_i^j b_j,~  
d_{i \otimes j}  =  c_i^j d_j,~ \\
u_{i \otimes j}  =  1-c_i^j(1-u_j),~
a_{i \otimes j}  = a_j, \nonumber 
\end{gather}
where $c_i^j$ is a discounting operator decided by different opinion models (detailed below), representing user $i$'s trust in user $j$. Finally, $\omega_i \oplus \omega_{i \otimes j}$ is obtained by: 
\begin{gather} \label{eq:consensus_op}
b_i \oplus b_{i \otimes j} =  \frac{b_i  (1-c_i^j(1-u_j)) + c_i^j b_j u_i}{\beta}~, \\ \nonumber
d_i \oplus d_{i \otimes j} =  \frac{d_i (1-c_i^j(1-u_j)) + c_i^j d_j u_i}{\beta}~, \\ \nonumber
u_i \oplus u_{i \otimes j} =  \frac{u_i(1 -c_i^j(1 - u_j))}{\beta}~, \\ \nonumber
a_i \oplus a_{i \otimes j} = \frac{(a_i - (a_i + a_j)u_i)(1 - c_i^j(1-u_j)) + a_j u_i}{\beta - u_i(1 - c_i^j(1-u_j))}~, \\ \nonumber
\beta = 1-c_i^j(1-u_i)(1-u_j)\neq 0,
\end{gather}
For simplicity, if uncertainty ($u$) is very low ($u \le T_u$), we assume $u=0$, halting further opinion updates by users.
 
We explore three opinion models (OMs) based on their types of trust, including uncertainty-based trust, homophily (like-mindedness)-based trust, and no-trust-based trust:
\begin{itemize}
\item {\bf Uncertainty-based OM (UOM)}: This OM uses uncertainty (or certainty)-based trust and calculates the uncertainty discounting operator $uc_i^j$ between two users as $uc_i^j = (1-u_i)(1-u_j)$. To ensure non-zero uncertainty for using the consensus operator in Eq.~\eqref{eq:consensus_op}, a {\em vacuity (uncertainty) maximization} technique~\cite{Josang16} is employed, defining a {\em vacuity-maximized opinion} for user $i$ as $\ddot{\omega}_i = (\ddot{b}_i, \ddot{d}_i, \ddot{u}_i, \bm{a}_i)$. Here, $\ddot{u}_i$ is the minimized {\em projected belief and disbelief}, adjusted by $a_i$ and $(1-a_i)$, with $\ddot{b}_i$ and $\ddot{d}_i$ calculated by reducing $\ddot{u}_i$ from $P (b_i)$ and $P(d_i)$, respectively.  When $u_i < \xi$ (indicating low uncertainty or vacuity with sufficient evidence) and dissonance (uncertainty caused by conflicting evidence) $\bm{b}_i^{\mathrm{Diss}} > T_d$ remains high, uncertainty $\ddot{u}_i$ is used to consider the effect of new information more dynamically. We set $\xi=0.01$ and $T_d = 0.6$ as thresholds to effectively maximize each party's influence based on our experimental analysis.

\item {\bf Homophily-based OM (HOM)}: Homophily, i.e., the tendency of like-minded individuals to associate, significantly influences opinion updates~\cite{Li11}. The homophily discounting operator $hc_i^j$ is determined using {\em cosine similarity}~\cite{Tan05} to measure the alignment between two users' beliefs and disbeliefs by:
\begin{eqnarray} \label{eq:hc}
hc_i^j & = & \frac{b_i b_j + d_i d_j }{\sqrt{b_i^2 + d_i^2} \sqrt{b_j^2 + d_j^2}}. 
\end{eqnarray}
Utilizing cosine similarity evaluates the similarity in opinions by focusing on belief and disbelief dimensions within an SL-based opinion model. This sidesteps direct consideration of uncertainty since belief and disbelief imply uncertainty due to $b+d+u=1$.
\item {\bf No-Trust-based OM (NOM)}: When we use neither UOM nor HOM, i.e., no trust filter used, we name the opinion model `No-Trust-based OM' and set $nc_i^j = 1$.
\end{itemize}

\subsubsection{\bf Opinion Reading}
Following~\cite{cho2019uncertainty}, we define a user's reading behavior as a probability $P_r$, indicating how often they read messages from neighbors. $P_r$ varies among 1 (multiple times per day), 0.5 (daily), 0.25 (weekly/monthly), and 0.1 (never) to model different user engagement levels with the received content. 

\subsubsection{\bf Opinion Sharing} \label{susbsubsec:opinion-sharing} Each user shares an opinion $\omega_i$ with friends $j$'s based on a sharing probability $P_s$. Following the distribution from~\cite{cho2019uncertainty}, $P_s$ varies to reflect sharing frequency: 1 (always/mostly), 0.5 (half the time), 0.25 (sometimes), and 0.1 (never). Considering reading messages motivates updating opinions~\cite{karnowski2018users}, we assume a user will possibly share only if they read before.

\subsection{Partially Observable Network} \label{subsec:uncertain-network-modeling}

Following~\cite{nasim2016PartialObservable}, we define a partially observable network within an undirected graph $G = (V, E)$. $G'$ = $(V, E')$ represents the known portion of the network, with $E' \subset E$. Network visibility probability $P_{nv}$ dictates the selection of $E'$, influencing the visibility of edges.

\subsection{DRL-based Seed Set Selection Process} \label{subsec:seed-set-selection}

We utilize DRL to guide the decision-making process, optimizing strategy selection for maximal influence spread within the network. We consider multi-round influence propagation. Each round consists of two steps: the false party (FP) selects one seed node and begins sharing assigned false information in the first step. After that, the true party (TP) follows in the second step with the assigned true information. An episode comprises a fixed number of rounds, which equals the number of seed nodes. The design of states, actions, and rewards of the DRL agent are described below.

\subsubsection{\bf States} At each step $t$, the state $s_t$ encompasses the network's current structure, defined as:
\begin{equation}
s_t = \Bigg \{\sum_{i,j \in \mathcal{U}_i} e_{i,j}, \max_{i \in \mathcal{U}_i} \mathrm{deg}_i \Bigg \},    
\end{equation}
where $\sum_{i,j \in \mathcal{U}_i} e_{i,j}$ calculates the number of edges among free nodes, and $\max_{i \in \mathcal{U}_i} \mathrm{deg}_i$ identifies the highest degree among these nodes. We define {\em free nodes} as $\{j| u_j^t \geq 0.5 \}$, meaning user $j$ still has a high uncertainty level and has yet to align with either party. Recall that $u_j^t$ refers to the `vacuity' (in Eq.~\eqref{eq:sl-mapping}) of user $j$ at step $t$.

\subsubsection{\bf Actions} \label{subsubsec: actions} Action space at step $t$ is ${\bf a}_t = \{a_t^{AF}, a_t^{BF}, a_t^{SGF}, a_t^{CF}\}$. Actions focus on the user's behavioral or opinion traits and their network position:
\begin{itemize}
\item \textit{Active First (AF, $a_t^{AU}$)} prioritizes the most active user, determined by the highest $P_r \times P_s$ in the network where $P_r$ and $P_s$ refer to a user's reading and sharing probabilities.
\item \textit{Blocking First (BF, $a_t^{B}$)} targets a user who is a neighbor of nodes belonging to the opponent's party and has the highest free degree (i.e., the maximum number of free nodes connected to it).  A user is considered to belong to a party based on its expected belief or disbelief. User $i$ belongs to the TP when $P(b_i) > 0.5$, or belongs to the FP if $P(d_i) > 0.5$.
\item \textit{SubGreedy First (SGF, $a_t^{SG}$)} chooses a node with the maximum number of neighboring nodes within $d$-hops~\cite{Lin15_LearningBasedCIM, Chung2019-DeepRL}.  This aims to efficiently capture a user's network power in a local network.  In this work, we choose $d=2$, which balances efficiency and effectiveness based on our experimental investigation. 

\item \textit{Centrality First (CF, $a_t^{CD}$)} selects a user with the highest degree centrality. 
\end{itemize}

\subsubsection{\bf Rewards} The reward at step $t$ by each party, $\mathcal{T}$ and $\mathcal{F}$, is estimated by:
\begin{equation}
\label{eq: instant reward}
R^\mathcal{T}_t = n^\mathcal{T}_t - n^\mathcal{T}_{t-2}, \; \; R^\mathcal{F}_t = n^\mathcal{F}_t - n^\mathcal{F}_{t-2}.
\end{equation}
The reward calculation for each party is based on the net change in the number of users aligned with them from the previous round. Specifically, for the FP which selects first, the reward at time $t=1$ is given by $R^\mathcal{F}_t = n^\mathcal{F}_t - n^\mathcal{F}_{t-1}$. The TP starts its calculation from $t=2$.  The accumulated reward in one episode is: 
\begin{equation}
\label{eq: accumulated reward}
R_T = \sum_{t=T}^{\infty}\gamma^{t-T+1}R_t, 
\end{equation}
where $R_t$ is either $R^\mathcal{T}_t$ or $R^\mathcal{F}_t$ in Eq.~\eqref{eq: instant reward} and $\gamma$ is a discounting factor, we choose $\gamma=0.95$ in this work.

\begin{figure}
    \centering
    \includegraphics[width= 0.48\textwidth, height = 0.33 \textwidth]{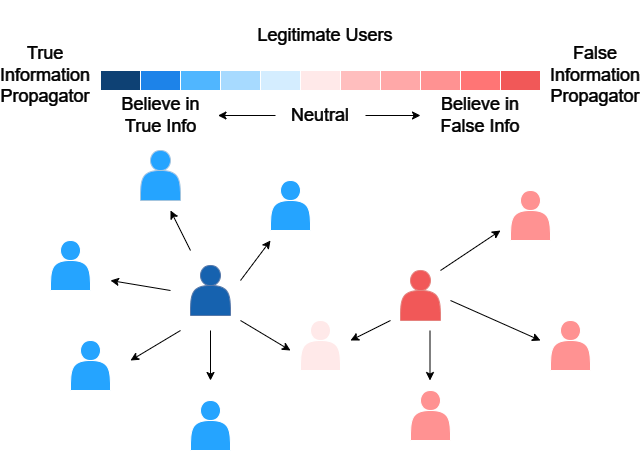}
    \caption{CIM using Subjective Logic-based opinion model}
    \label{fig:cim_overview}
    \vspace{-5mm}
\end{figure}

Fig.~\ref{fig:cim_overview} presents an overview of the CIM problem using Subjective Logic, illustrated through an OSN case example. The figure employs a color gradient to represent belief strength: lighter colors indicate less certainty, with white denoting neutral users. Blue signifies users inclined toward true information, with the darkest blue representing true information propagators. Conversely, red indicates users leaning towards false information, with the darkest red marking FIP identified by the FP.

\section{Experiment Setup} \label{sec:exp-setup}

{\bf Experimental Setting:}  
We employ Proximal Policy Optimization (PPO)~\cite{schulman2017ppo} as a DRL algorithm to select optimal seed nodes because PPO allows agents to update multiple times with the same group of data with reduced computational complexity.  We assume six possible FPs using different strategies, including the four strategies in the action space (i.e., AF, BF, CF, SGF), Random, and DRL. `Random' means randomly choosing a strategy out of the action space. We train six distinct DRL agents for TP, given each FP, and then test with the corresponding FP in Section~\ref{sec:numerical-analysis-results}. In reality, true information often counters false narratives, so we give FIPs the first-mover advantage to select seed nodes and propagate information {\em first} in each round using {\em Breadth-First Search} for information spread.  We initialize SL-based opinions for legitimate users using the mapping rule in Eq.~\eqref{eq:sl-mapping} with $(r, s, W) = (1, 1, 101)$, indicating high uncertainty due to insufficient evidence.  For TIPs, seed nodes' opinions are set to $(r, s, W) = (100, 1, 2)$, denoting strong confidence in their true information. Conversely, FIPs have opinions initialized at $(r, s, W) = (1, 100, 2)$, reflecting confidence in the false information.
We use an HPE Apollo 6500 system equipped with AMD EPYC 7742 chips, featuring a base frequency of 2.25 GHz and a boost of up to 3.4 GHz for all experiments.

{\bf Datasets:} We use the URV Email Network~\cite{rossi2015network_dataset_c} for the current experimental results. The dataset is the email communications at {\em Universitat Rovira i Virgili in Spain}, an undirected graph with 1,133 nodes and 5,452 edges.

{\bf Metrics:} The effectiveness of each party's information maximization is measured by the number of users holding the party's beliefs, as determined by Eq.~\eqref{eq:expected-opinion}. Users with $P_i(b_i) \geq 0.5$ are classified as TP, indicating TP's influence, denoted by $n^T$. Algorithmic efficiency is assessed by the simulation's running time per episode.

{\bf Comparing Schemes:} \label{subsec:comparing-schemes} Unlike~\cite{Bharathi07-CIM, Chung2019-DeepRL, Lin15_LearningBasedCIM, Pham19-CIM}, our approach employs a dynamic opinion model, allowing user opinions to evolve through interactions using SL~\cite{Josang16}. To ensure a fair comparison with existing state-of-the-art (SOTA) approaches, all models will operate under the same conditions. This includes utilizing the same opinion models (HOM, UOM, or NOM), identical seed node selection counts (i.e., $k=50$), and opinion updating mechanisms as detailed in Section~\ref{sec:our-algorithm}.  Adapting these schemes to our settings might alter their original performance, but this standardization is crucial for assessing their performance in the more realistic scenarios addressed in our study. We evaluate the performance of the following CIM algorithms:
\begin{itemize}
\item {\bf DRIM-A} implements a complete action set for seed node selection as detailed in Section~\ref{subsec:seed-set-selection}. `A' represents the AF strategy. 
\item {\bf DRIM-NA} utilizes a subset of actions, specifically excluding AF where NA represents No-AF, for seed node selection.
\item {\bf STORM} is a SOTA approach in~\cite{Lin15_LearningBasedCIM}. For a fair comparison, we apply our user opinion model to STORM where we define free node ($\{j| u_j^t \geq 0.5 \}$) as the non-occupied node, which is different from the concept of free nodes used in~\cite{Lin15_LearningBasedCIM}.  We also simplify STORM's action space as max-weight and max-degree are the same in an unweighted graph.  
\item {\bf C-STORM}~\cite{Chung2019-DeepRL} advances STORM by adding a preliminary step of identifying the optimal {\em community} for seed selection detected by a community detection algorithm, named it as C-STORM (i.e., community-based STORM). For a fair comparison, we use our opinion model and the free nodes defined in our work. 
\end{itemize}

\section{Numerical Analysis \& Results} \label{sec:numerical-analysis-results}

We apply the trained model as the node selection agent when applying the DRL-based strategy. Then, we let both parties pick 50 seeds and propagate in turns. We assume TIP propagates twice after picking a seed while FIP does once. We make this assumption because a mix of human judgment, diversity in algorithm design, and public awareness play critical roles in the ongoing battle against false information on OSNs, it is natural that true information can gain more valid propagation in OSNs. All the experiment results have an average of 20 runs.

\subsection{Effect of Various Opinion Models} \label{subsec:OM-Vary}

\begin{table}
\centering
\caption{\sc \centering True Party (TP)'s Influence in $n^\mathcal{T}$ Under Various CIM Algorithms with Different OMs}
\label{tab:OM-vary-table}
\vspace{-2mm}
\scriptsize
\begin{tabular}{|P{1.8cm}|P{0.8cm}|P{0.4cm}|P{0.5cm}|P{0.5cm}|P{0.4cm}|P{0.5cm}|}
\hline
\textbf{Scheme / OM} & \textbf{Random} & \textbf{AF} & \textbf{BF} & \textbf{SGF} & \textbf{CF} & \textbf{DRL} \\ \hline
DRIM-A/UOM & \textbf{937.5} & \textbf{1008}  & \textbf{894.3} & \textbf{951.4} & \textbf{937.3} & \textbf{890.8} \\
DRIM-A/HOM & 150.5 & 61.5  & 576.3 & 57    & 227.6 & 142.9 \\ 
DRIM-A/NOM & 474.9 & 89    & 561.3 & 60.2  & 65.1  & 102.2 \\  \hline
DRIM-NA/UOM & \textbf{897.3} & \textbf{992.4} & \textbf{826.8} & \textbf{937.9} & \textbf{808.6} & \textbf{926.9} \\ 
DRIM-NA/HOM & 232.6 & 48.2  & 672.2 & 109.3 & 62.9  & 53.6  \\  
DRIM-NA/NOM & 476.2 & 79.8  & 486.8 & 66.6  & 76.2  & 150.5 \\ \hline
C-STORM/UOM & \textbf{659.7} & \textbf{704.2} & \textbf{1051.3} & \textbf{858.3} & \textbf{898} & \textbf{743} \\ 
C-STORM/HOM & 238.3 & 618  & 679.7 & 149.8 & 64.2 & 58.7 \\  
C-STORM/NOM & 563.2 & 586.7  & 403.8 & 74.4 & 77.1 & 65.7 \\ \hline
STORM/UOM & \textbf{722.3} & 652 & \textbf{1007.3} & \textbf{861} & \textbf{877} & \textbf{674.7} \\ 
STORM/HOM & 373.7 & \textbf{659.2} & 779.3 & 62.4 & 62.1  & 68.6  \\  
STORM/NOM & 341 & 477.6  & 659.1 & 74.4 & 72.7 & 358.3 \\ \hline
\end{tabular}
\vspace{-5mm}
\end{table}


We compare the performance of four schemes against various-strategy FP under three different opinion models: UOM, HOM, and NOM. As shown in Table~\ref{tab:OM-vary-table}, the first column marks the scheme names associated with the opinion models, and the first row represents the node selection strategy applied by FP. Performance is represented by the number of nodes $n^\mathcal{T}$ aligned with TP $\mathcal{T}$. As shown in Table~\ref{tab:OM-vary-table}, UOM always lets true information gain the highest impact. This is because when user $i$ adopts UOM, user $i$ tends to accept new evidence when the $u_i$ is low. Thus, even with the first-mover advantage of FP, true information can still be well spread through the OSN.

The poor performance with HOM and NOM also shows that it is crucial to reach out to the users with the correct information earlier rather than later. It requires fast detection and quick response to combat False information.  

\subsection{Effect of DRL-based TP's Influence Under Various Strategies Taken by FP} \label{subsec:T-DRL-F-Vary}

\begin{figure}
    \centering
        \subfigure{     \includegraphics[width=0.45\textwidth, height=0.025\textwidth]{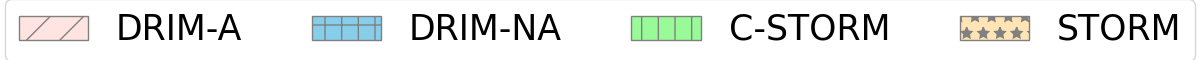}}
    \setcounter{subfigure}{0}
    \includegraphics[width=0.4\textwidth]{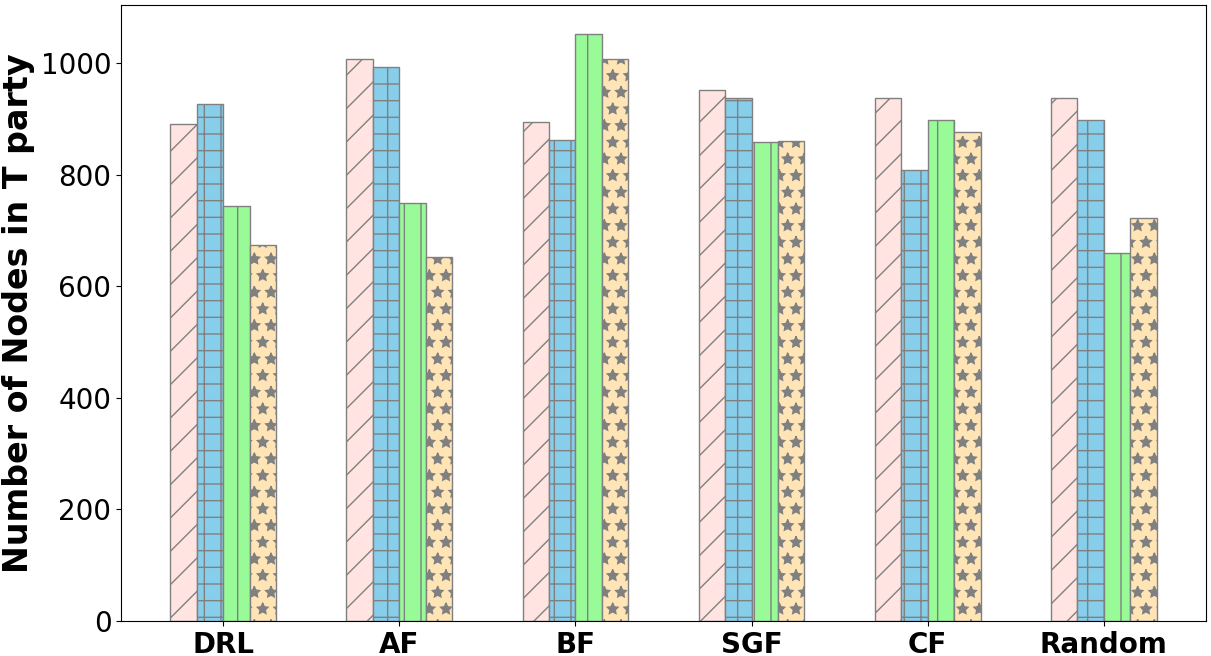}
    \caption{TP's influence under various CIM algorithms}
    \label{fig:T-drl-F-vary-strategy}
    \vspace{-5mm}
\end{figure}

\begin{figure*}[htb]
  \centering
  \subfigure{
    \includegraphics[width=0.55\textwidth, height=0.025\textwidth]{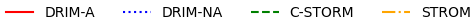}}
    \setcounter{subfigure}{0}
    \vspace{-3mm}    
    
  \subfigure[Varying \# of information propagation (IP)]{
    \includegraphics[width=0.31\textwidth, height=0.25\textwidth]{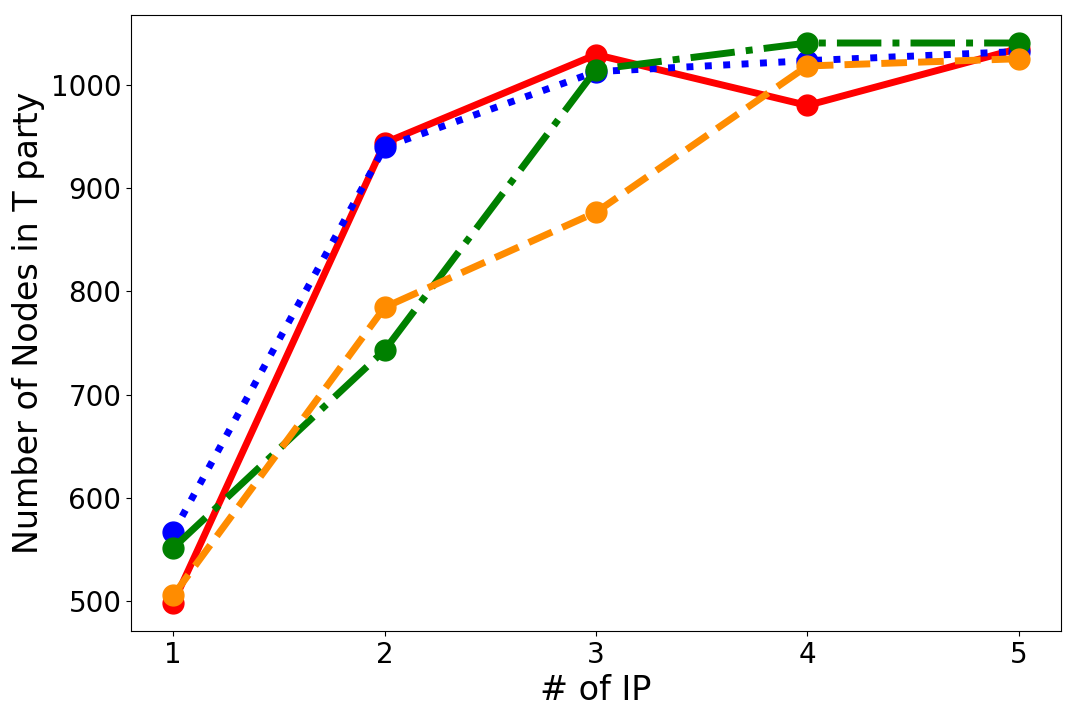}\label{fig:drl-drl-IP-sens}}
  \subfigure[Varying \% of network observability]{
    \includegraphics[width=0.31\textwidth, height=0.25\textwidth]{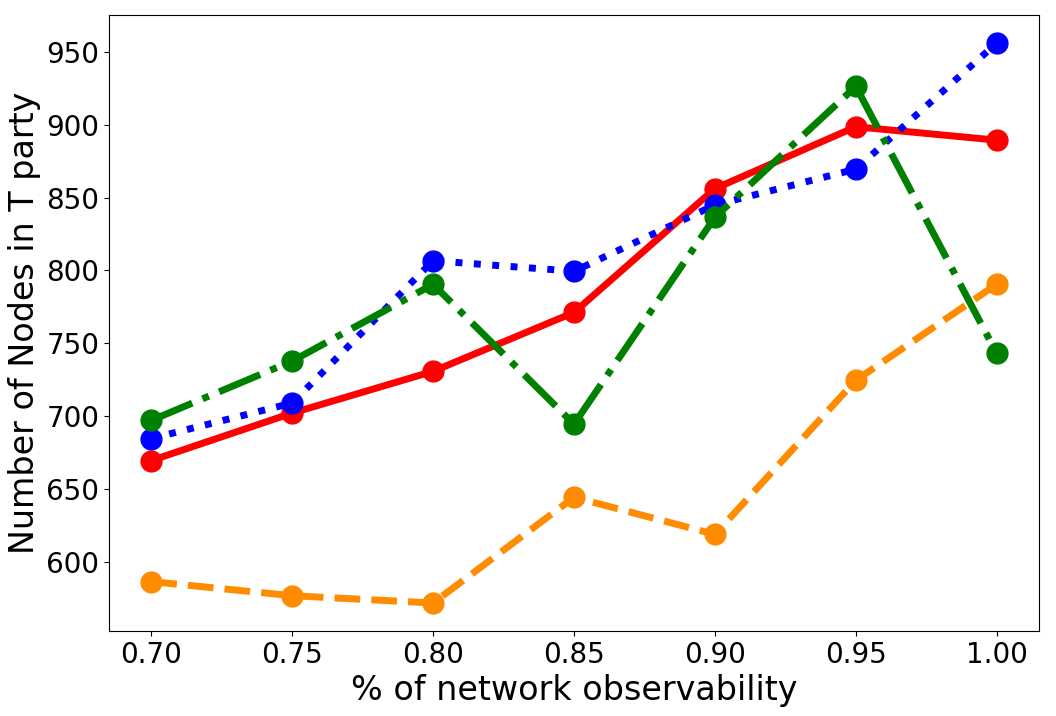} \label{fig:drl-drl-PON-sens}}
  \subfigure[Varying users' prior belief]{
    \includegraphics[width=0.31\textwidth, height=0.25\textwidth]{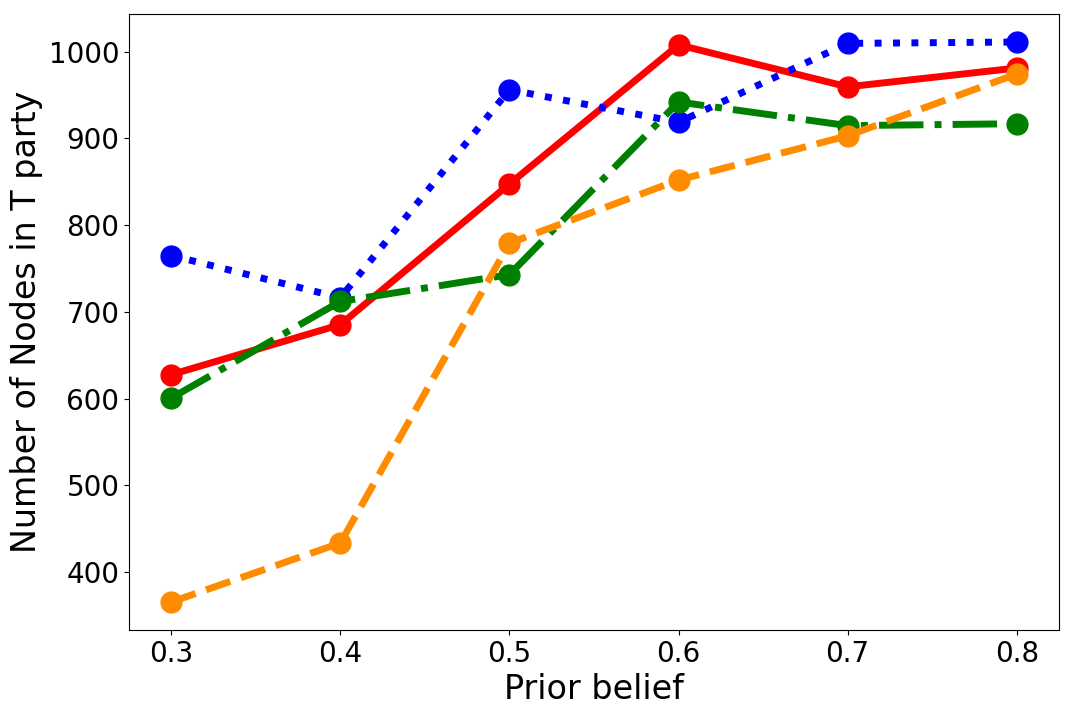} \label{fig:drl-drl-base-rate-sens}}
    \vspace{-2mm}
    
    \caption{True party (TP)'s influence under various CIM algorithms when both TP and False party (FP) use DRL for seed node selection}
\label{fig:drl-sens}
\vspace{-6mm}
\end{figure*}

Fig.~\ref{fig:T-drl-F-vary-strategy}
presents the comparative performance analysis of the TP's influence under various CIM algorithms and UOM when the TP selects strategies to find seed nodes using DRL. At the same time, the FP takes the fixed strategy to select seed nodes under a fully observable network. It illustrates the effectiveness of these four CIM algorithms under UOM when the FP chooses seed nodes by a random strategy, one of four strategies in the action space, or DRL-agent.   Except for Blocking First (BF, $a_t^{B}$), when the FP applies the other five strategies, DRIM-A and DRIM-NA outperform C-STORM and STORM. The reason for lower performance under DRIM-A and DRIM-NA with the opponent taking BF is that blocking can mitigate the impact of opponents but is not effective in increasing its party's influence. Therefore, our schemes are more suitable for proactive opponents.

DRIM-A outperforms the DRIM-NA in most of the cases as the beneficial of Active First (AF, $a_t^{AU}$) action. The gap is distinct, especially when FP uses CF, a common and widely accepted strategy in IM. Hence, the user behavior is worth considering in the CIM problem.

\subsection{Sensitivity Analyses}
This section discusses the sensitivity analyses of how the number of times propagating true information or the percentage of network observability affects the performance of the considered CIM algorithms that use UOM as an opinion model.

\subsubsection{Effect of Varying the Number of Information Propagation by TIPs}
In Fig.~\ref{fig:drl-drl-IP-sens}, we examine the effect of increasing the number of information propagation (IP) instances for the TP from 1 to 5 while the FP maintains a fixed, single strategy. Naturally, all schemes demonstrate increased influence with more IPs. Specifically, our schemes, DRIM-A and DRIM-NA, exhibit a significant gain in influence when IP is increased from once to twice per step, efficiently countering false information with minimal additional resource investment. Beyond IP $> 2$, the increase in influence plateaus, which justifies setting the default IP value at 2 to optimize resource use while maximizing influence.

\subsubsection{Effect of Varying the Degree of Network Observability}
As depicted in Fig.~\ref{fig:drl-drl-PON-sens}, DRIM-A and DRIM-NA maintain influence comparable to C-STORM under conditions of partial network observability, impacting both the TP and FP. This shared visibility condition means that any increase in network transparency benefits FP's performance as well. In contrast, STORM and C-STORM show fluctuating performance improvements with greater network visibility, indicating their lack of robustness against opponents who possess enhanced information.

\subsubsection{Effect of Varying the users' prior belief}

Prior belief represents the initial probability favoring belief acceptance. A higher prior belief enhances TP's influence, as demonstrated in Fig.~\ref{fig:drl-drl-base-rate-sens}. Notably, even with low ($a<0.5$) or neutral ($a=0.5$) prior belief, DRIM-A and DRIM-NA outperform C-STORM and STORM, indicating effectiveness even when initial user beliefs do not favor true information.  In summary, DRIM-A and DRIM-NA are resource-efficient and robust in partially observable networks. They effectively combat false information by maximizing the impact of true information, even when users initially lean towards false information.

\begin{table}[t]
\centering
\caption{\sc \centering Simulation Running Time (RT) of the Considered CIM Algorithms}
\vspace{-2mm}
\footnotesize 
\begin{tabular}{|P{1.1cm}|P{1.2cm}|P{1.4cm}|P{1.4cm}|P{1cm}|}
\hline
Algo. & DRIM-A & DRIM-NA & C-STORM & STORM \\
\hline
RT (sec.) & 0.108 & 0.099 & 0.132 & 0.244\\
\hline
\end{tabular}
\label{tab:running_time}
\vspace{-5mm}
\end{table}

\subsection{Running Time Analysis of CIM Algorithms} 
\label{subsec:complexity-analysis}
Our DRIM-A and DRIM-NA demonstrated superior running time as shown in Table~\ref{tab:running_time}. DRIM-NA is more efficient because of the smaller action space, but sacrificing the performance as shown in Fig.~\ref{fig:T-drl-F-vary-strategy}.

\section{Conclusions \& Future Work} \label{sec:conclusions-future-work}

This work introduced a deep reinforcement learning (DRL)-based framework enhanced with Subjective Logic (SL) to refine competitive influence maximization (CIM) analysis by integrating uncertain opinions and user preferences in Online Social Networks (OSNs). Unlike traditional CIM approaches that rely on binary opinions (i.e., either 0 or 1), our model incorporates a more nuanced Uncertainty-based Opinion Model (UOM), which reflects a more realistic representation of user attitudes. Our proposed schemes, DRIM-A and DRIM-NA, demonstrate superior efficiency and effectiveness in countering false information by leveraging this model.

Our experimental results demonstrated that UOM significantly enhances the dissemination of true information, even with less engaged users, a typical scenario in combating false information. Further, recognizing user behavior patterns and engagement levels is crucial when designing the action space for the True Party, particularly as active users with high reading and sharing frequency.

{\bf Future work} aims to extend these experiments to larger and denser social networks to evaluate the scalability of our framework. We will explore the nuances of uncertainty further to develop more sophisticated and practical solutions, contributing significantly to advancing CIM research.

\section*{Acknowledgement}
This work is partly supported by NSF under Grant Numbers 2107449, 2107450, and 2107451. The views and conclusions contained in this document are those of the authors and should not be interpreted as representing the official policies, expressed or implied, of the U.S. Government. 

\bibliographystyle{ACM-Reference-Format}
\bibliography{ref}

\end{document}